\documentclass[12pt]{article}
\usepackage{epsfig}

\newcommand{\agt}{\,\rlap{\lower 3.5 pt \hbox{$\mathchar \sim$}} \raise 1pt
 \hbox {$>$}\,}
\newcommand{\alt}{\,\rlap{\lower 3.5 pt \hbox{$\mathchar \sim$}} \raise 1pt
 \hbox {$<$}\,}

\begin{document}

\title{
\vskip-3cm{\baselineskip14pt
\centerline{\normalsize DESY 10-026\hfill ISSN 0418-9833}
\centerline{\normalsize March 2010\hfill}}
\vskip1.5cm
\bf Factorization breaking in high-transverse-momentum charged-hadron
production at the Tevatron?}

\author{S. Albino, B.~A. Kniehl, G. Kramer
\\
{\normalsize\it II. Institut f\"ur Theoretische Physik, Universit\"at
Hamburg,}
\\
{\normalsize\it Luruper Chaussee 149, 22761 Hamburg, Germany.}}

\date{}

\maketitle

\begin{abstract}
We compare the transverse momentum ($p_T$) distribution of inclusive
light-charged-particle production measured by the CDF Collaboration at the
Fermilab Tevatron with the theoretical prediction evaluated at next-to-leading
order in quantum chromodynamics (QCD) using fragmentation functions recently
determined through a global data fit.
While, in the lower $p_T$ range, the data agree with the prediction within the
theoretical error or slightly undershoot it, they significantly exceed it in
the upper $p_T$ range, by several orders of magnitude at the largest values of
$p_T$, where perturbation theory should be most reliable.
This disagreement is too large to be remedied by introducing additional
produced particles into the calculation, and potentially challenges the
validity of the factorization theorem on which the parton model of QCD relies.
Clearly, a breakdown of the factorization theorem, being a fundamental property
of QCD, would be extremely difficult to understand.

\medskip

\end{abstract}

\newpage

Inclusive single-hadron production processes in $e^+ e^-$, $ep$, and
$pp(\bar{p})$ reactions can be well described
\cite{Kniehl:2000hk,Albino:2006wz} by cross section calculations that depend on
universal fragmentation functions (FFs), consistent with the validity of the
factorization theorem at CERN LEP, BNL RHIC, and DESY HERA energies (for a
recent review, see Ref.~\cite{Albino:2008gy}).
However, in order to prepare for incoming data at much higher energy from the
CERN LHC, from the late 2009 run and/or from future runs, this validity needs
to be tested also at the Tevatron.
Such tests were performed in Ref.~\cite{Kniehl:2000hk}, using CDF data for
which $p_T$ reached 10~GeV \cite{Abe:1988yu} and also UA1 and UA2 data from
CERN S$p\bar p$S runs at sub-TeV energies.  
Very recently, new data have been obtained by the CDF Collaboration
\cite{Aaltonen:2009ne} for which $p_T$ reaches 150~GeV. 
Such data provide a unique opportunity to test the factorization theorem
applied to the final state in a high-$p_T$ regime of inclusive hadron
production that has never been accessible before, where we expect perturbation
theory in the QCD-improved parton model to work perfectly, and these data are
the focus of this Letter.
We note that agreement between theory and recent CDF measurements of inclusive
jet production \cite{Aaltonen:2008eq} has been obtained at much higher $p_T$
values, which confirms the inital-state factorization theorem for these data. 
Such measurements originate from the same physical processes as the
lower-$p_T$ hadron production measurements do, since the produced hadron
carries just a fraction of the jet's energy.

We present here the first rigorous interpretation of the CDF data of
Ref.~\cite{Aaltonen:2009ne} in the context of perturbative QCD.
The experimental analysis resorted to the Monte Carlo event generator PYTHIA,
which failed to give useful results for $p_T > 50$~GeV.
We use the latest FF sets \cite{Hirai:2007cx,deFlorian:2007aj,Albino:2008fy}
extracted from global data fits that included accurate primary-quark-tagged
measurements from $e^+ e^-$ reactions.
Unfortunately, $e^+ e^-$ reaction data do not sufficiently constrain the gluon
FF, because it only enters the calculation at next-to-leading order (NLO),
while it enters the calculation of $p\bar{p}$ reactions at the same order as
the quark FFs.
The two most recent FF sets, AKK08 \cite{Albino:2008fy} and DSS
\cite{deFlorian:2007aj}, also used $pp$ reaction data from RHIC, but the gluon
FFs in both cases differ substantially, suggesting that the RHIC data used in
those fits did not significantly improve the constraints on the gluon FFs.
The FF errors in both sets were not determined.
Therefore, the best estimate of the error on the predictions for CDF data at
present is given by the general spread in the calculation using different sets.
Besides the AKK08 and DSS FF sets, we also include the HKNS one
\cite{Hirai:2007cx}, which was obtained from a fit only to $e^+ e^-$ data.
We assume that the unidentified charged particle in these CDF measurements is
a light charged hadron.
That is, we take the sum of the $\pi^\pm$, $K^\pm$ and $p/\bar{p}$ production
cross sections, and assume the possible contamination with other charged
particles to be negligible.

Our calculations are performed to NLO in the modified minimal-subtraction
($\overline{\rm MS}$) renormalization and factorization schemes.
We set the renormalization scale to be $\mu=k p_T$ and the factorization scale
to be $\mu_f=k_f p_T$, where $k=k_f=1$ unless otherwise stated. 
For the initial protons, we use the parton density function (PDF) set
CTEQ6.6M \cite{Nadolsky:2008zw}.
Since the produced-hadron mass was accounted for only in the AKK08 analysis,
we account for it in the same way here for calculations using the AKK08 FF
sets, but set it to zero when using the DSS and HKNS sets.
In any case, neglecting the hadron masses makes very little difference, as we
shall see later.
We refer to the above calculation as the AKK08 default one (labeled
``AKK08(default)'' in our figures).

\begin{figure*}[t]
\begin{center}
\includegraphics[width=15.cm]{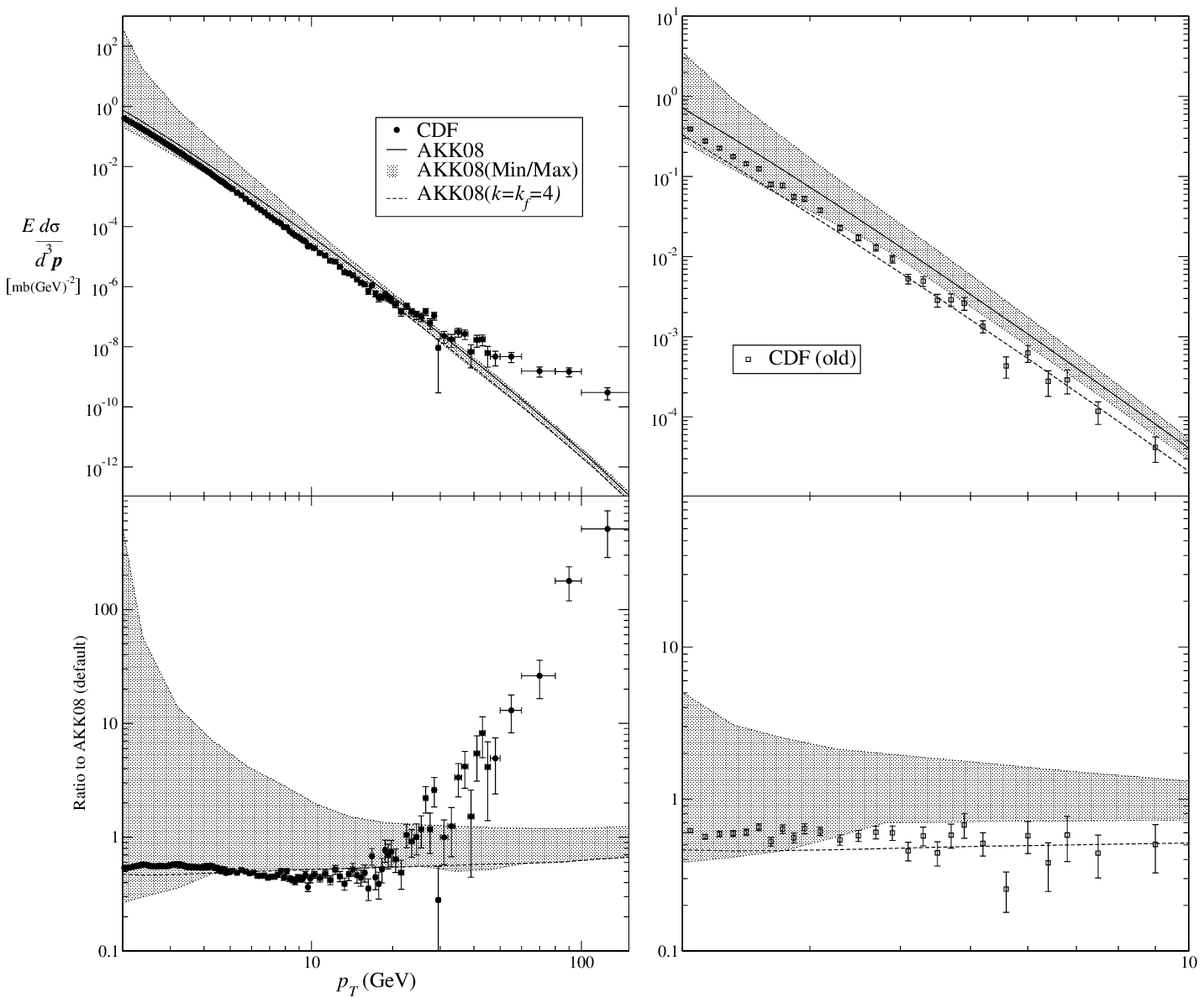}
\end{center}
\vspace{-0.5cm}
\caption{{\bf Top left}: The CDF data of Ref.~\cite{Aaltonen:2009ne} for
inclusive unidentified-charged-particle production is compared with NLO QCD
predictions evaluated using the AKK08 FF sets \cite{Albino:2008fy}.
The shaded region labeled ``AKK08(Min/Max)'' indicates, for each $p_T$ value,
the maximum and minimum values of the cross section within the ranges
$1/2 < k <2$ and $1/2 < k_f <2$.
Also shown is the result for $k=k_f = 4$.
{\bf Top right}: As in the top left plot, except for the older CDF data of
Ref.~\cite{Abe:1988yu}.
Each lower plot is identical to the one above it, except that all results are
divided by the central NLO predictions for clarity.
(The latter hence correspond to unity.)
\label{CDF1}}
\end{figure*}

The comparison with the recent data \cite{Aaltonen:2009ne} collected in
Run~II (with center-of-mass energy $\sqrt{s}=1.96$~TeV) in the central region
of the CDF detector (with pseudorapidity $|\eta|<1$) and older CDF data
\cite{Abe:1988yu} from Run~I ($\sqrt{s}=1.8$~TeV, $|\eta|<1$) is shown in
Fig.~\ref{CDF1}.
The most striking observation to be made is the disagreement at $p_T>50$~GeV
(see left plot).
Looking closely at the experimental line shape, there appears to be a
transition to a softer slope, corresponding to a fall-off with a lower power
in $1/p_T$, at about $p_T=20$~GeV, which eventually leads to a systematic
departure from the theoretical prediction.
In fact, fitting the function $A/p_T^n$ to the CDF data yields $A=71.50$ and
$n=6.52$ with $\chi^2/{\rm d.o.f.}=0.886$ for 5~GeV${}<p_T<20$~GeV and
$A=0.030$ and $n=3.97$ with $\chi^2/{\rm d.o.f.}=2.18$ for
20~GeV${}<p_T<150$~GeV.
We evaluate the theoretical error, indicated by the shaded region labeled
``AKK08(Min/Max)'' in Fig.~\ref{CDF1} and later plots, by plotting, for each
$p_T$ value, the maximum and minimum cross sections in the range
$1/2 < k,k_f < 2$.
To achieve this, we use a 17$\times$17 grid of points equidistant in
$\ln k_{(f)}$ ranging from $-\ln 2$ to $\ln 2$.
As Fig.~\ref{CDF1} shows, the theoretical error calculated in this way is not
nearly sufficient to accommodate the recent CDF data at large $p_T$.
Within this error, the calculation in the region $p_T \alt 40$~GeV is
consistent with most of the data.
In the region $p_T \alt 20$~GeV, the data lie in the lower range of the
theoretical error, and in fact favour a large value of $k=k_f$, as can be seen
by the dashed curve in Fig.~\ref{CDF1}, which corresponds to $k=k_f=4$.
The error is largest at small $p_T={\mathcal O}(1)$, where both
$\alpha_s(\mu)$ and the partonic cross sections become very large, and for
this reason we limit our discussion to those data for which $p_T >2$~GeV.

\begin{figure}[h!]
\begin{center}
\includegraphics[width=6cm,angle=-90,bb=80 10 580 710,clip]{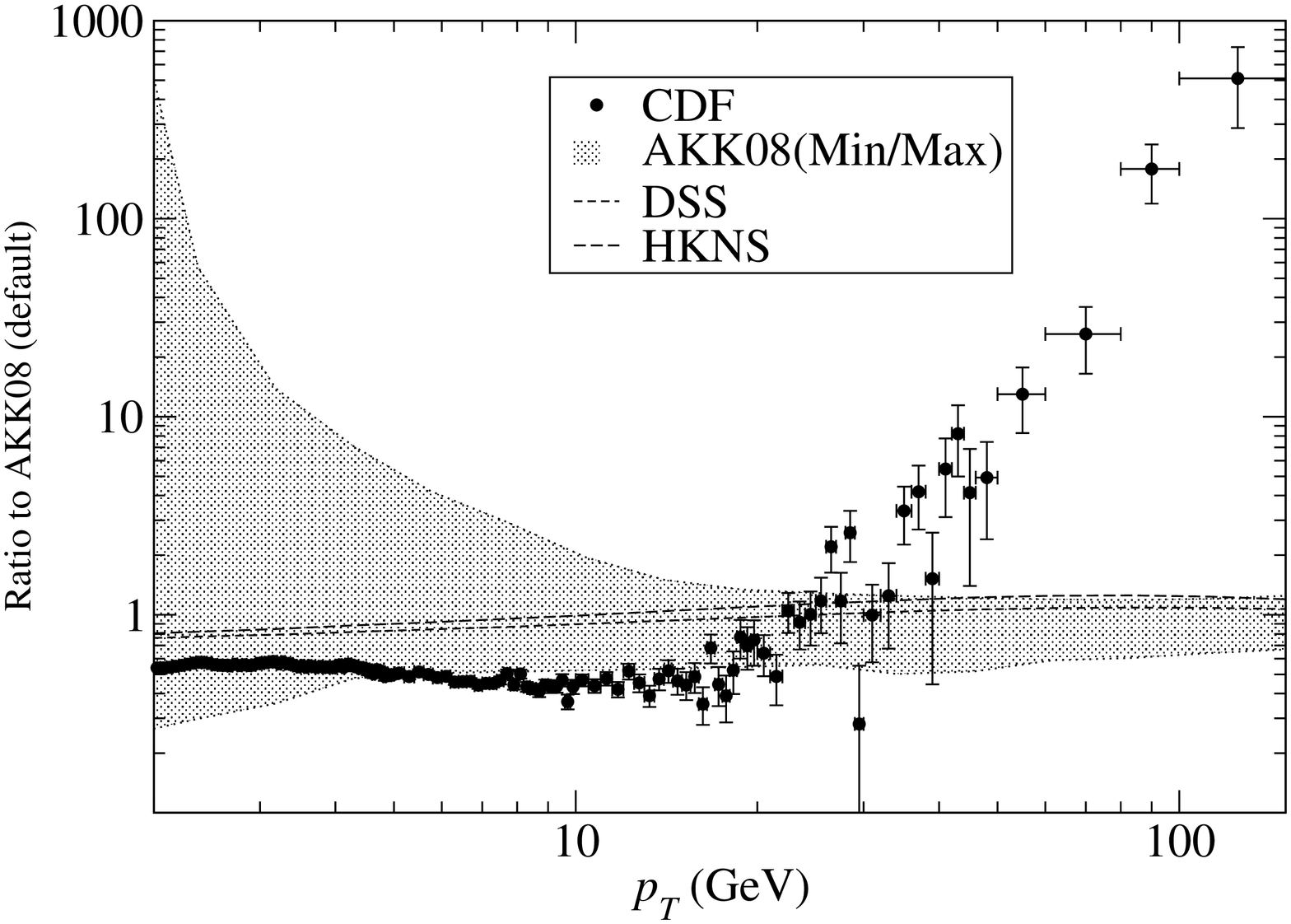}
\end{center}
\vspace{-0.5cm}
\caption{As in Fig.~\ref{CDF1} (bottom left).
Also shown is the calculation performed using the DSS \cite{de Florian:2007hc}
and HKNS \cite{Hirai:2007cx} FF sets.
\label{CDF2}}
\end{figure}

In Fig.~\ref{CDF2}, the same calculation is performed with the DSS
\cite{de Florian:2007hc} and HKNS \cite{Hirai:2007cx} FF sets.
We note that FFs for unidentified charged hadrons were also obtained by the
authors of Ref.~\cite{de Florian:2007hc}.
However, the difference between the calculation using those FFs and the one
using the sum of the DSS FF sets for $\pi^\pm$, $K^\pm$, and $p/\bar{p}$ 
production (which respectively make up about 75, 15, and 10\% of the sample) 
turns out to be negligible over the whole $p_T$ range considered.
Assuming that the difference between the calculations using the AKK08, DSS,
and HKNS FF sets gives an estimate of the error due to the uncertainties in
the FFs, we conclude that the latter are not nearly sufficient to account for
any discrepancy between theory and the new CDF data \cite{Aaltonen:2009ne}.
We stress that all these FF sets are tightly constrained in the very region of
the $(z,\mu_f)$ parameter space that is probed by our predictions, in
particular where the discrepancy occurs, so that no extrapolation error arises.
In fact, the average value of $z$ lies in the range
$0.44 \alt \langle z \rangle \alt 0.65$ for
2~GeV${}<\mu_f=p_T<150$~GeV.
The wealth of data from TRISTAN, LEP1, SLC, and LEP2, which populate the range
58~GeV${}<\mu_f=\sqrt{s}<191$~GeV, are quite confining in this intermediate
$z$ range. 
We conclude that the error due to the uncertainties in the FFs is small
against the overall one.

\begin{figure}[h!]
\begin{center}
\includegraphics[width=6cm,angle=-90,bb=80 10 580 710,clip]{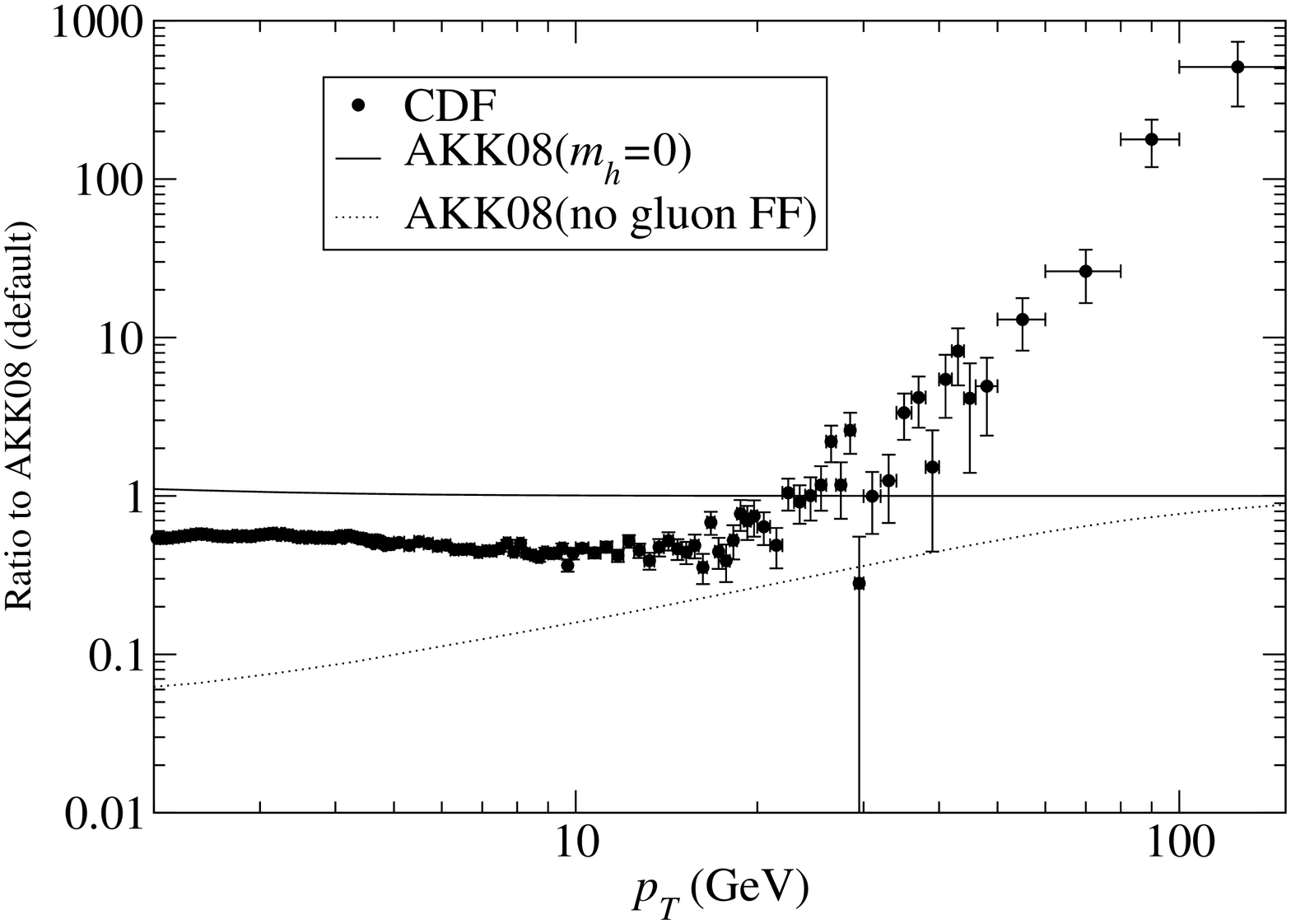}
\end{center}
\vspace{-0.5cm}
\caption{As in Fig.~\ref{CDF1} (bottom left) for the data.
The AKK08 calculation is performed with all hadrons' masses set to zero
(labeled ``$m_h=0$'') and again without the contribution from the evolved gluon
FF (``no gluon''). \label{CDF3}}
\end{figure}

In Fig.~\ref{CDF3}, we investigate the effect of the produced hadron's mass
$m_h$, by comparing our default prediction with the one obtained by setting
all particles' masses to zero. 
The difference also gives an estimate for the size of any neglected
small-$p_T$ effect of a similar order of magnitude, namely $O(m_h^2/p_T^2)$,
such as higher twist.
Clearly, such effects are negligible relative to the theoretical uncertainty
shown in Fig.~\ref{CDF1}.

As mentioned above, the gluon FF is somewhat less well constrained than the
quark FFs.
In order to determine whether the discrepancy with the recent CDF data
\cite{Aaltonen:2009ne} could be due to large errors on the gluon FF, we also
show in Fig.~\ref{CDF3} the contribution coming from quark fragmentation only
(labeled ``no gluon'').
We conclude that a reasonable modification to the gluon FF may improve the
description of the low-$p_T$ data, but certainly not at the largest $p_T$
values where gluon fragmentation becomes negligible.

\begin{figure}[h!]
\begin{center}
\includegraphics[width=6cm,angle=-90,bb=80 50 580 710,clip]{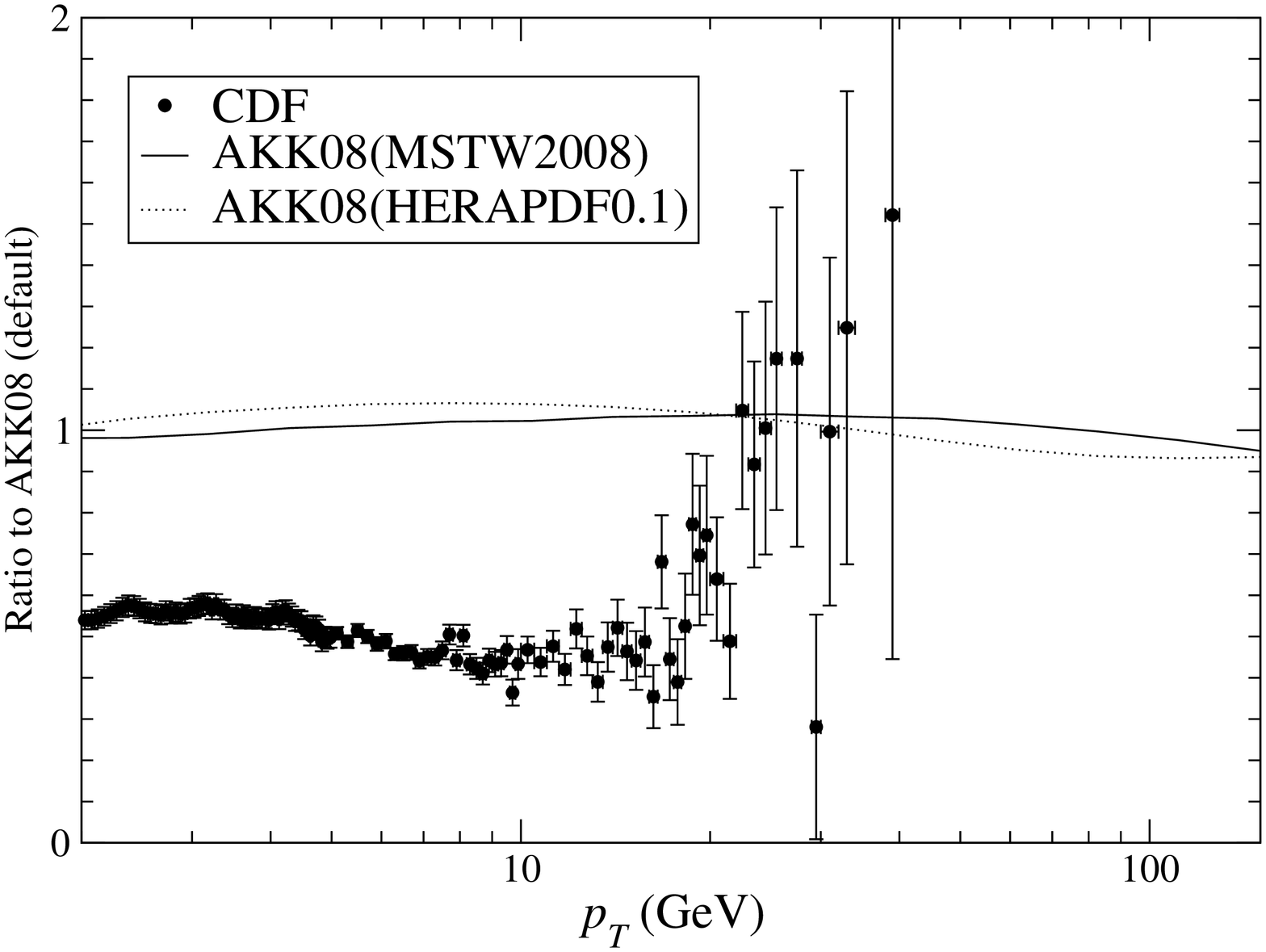}
\end{center}
\vspace{-0.5cm}
\caption{As in Fig.~\ref{CDF1} (bottom left) for the data.
The AKK08 calculation is performed using the PDF sets of
Ref.~\cite{Martin:2009iq} (labeled ``MSTW'') and Ref.~\cite{HERAPDF}
(labeled ``HERAPDF0.1'').
The data for which $p_T>50$~GeV lie far above the calculation.
\label{CDF6}}
\end{figure}

The only other non-perturbative quantities in our analysis besides FFs are
PDFs.
Also these are tightly constrained by global analyses in the range of $x$ and
$\mu_f$ probed by the CDF data.
In order to illustrate this, we study in Fig.~\ref{CDF6} the variation due to
the choice of PDF set, by comparing our default result based on the CTEQ 6.6M
PDF set \cite{Nadolsky:2008zw} with the calculations using the MSTW2008
\cite{Martin:2009iq} and HERAPDF0.1 \cite{HERAPDF} ones.
All these PDF sets have been constrained in very different ways.
Specifically, the HERAPDF0.1 PDF set \cite{HERAPDF} was obtained from a
combined fit to HERA data, which puts strong constraints on the gluon PDF at
low $x$.
Clearly this type of error is the least significant of all.

\begin{figure}[t!]
\begin{center}
\includegraphics[width=6cm,angle=-90,bb=80 20 580 720,clip]{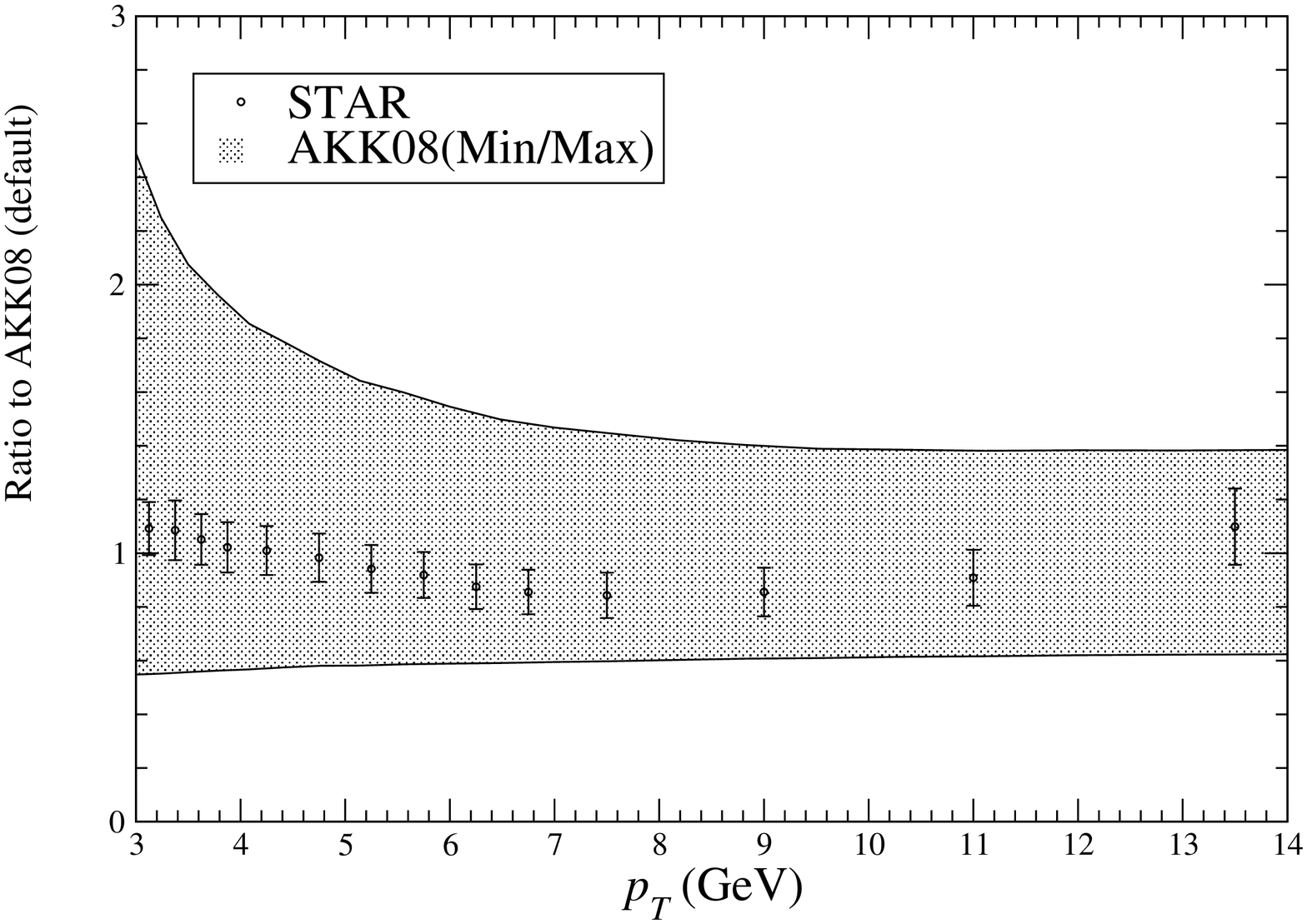}
\end{center}
\vspace{-0.5cm}
\caption{The STAR data of Ref.~\cite{Xu:2009aa} for inclusive charged-hadron
production is compared with NLO QCD predictions evaluated using the AKK08 FF
sets \cite{Albino:2008fy}.\label{STAR1_5}}
\end{figure}

Finally, we compare with the recent accurate data from STAR \cite{Xu:2009aa}
($\sqrt{s}=200$~GeV, $|y|<0.5$) in Fig.~\ref{STAR1_5}, and find perfect
agreement with our default prediction, well within the theoretical error.
By contrast, the CDF data from Runs I \cite{Abe:1988yu} and II
\cite{Aaltonen:2009ne} tend to undershoot the theoretical predictions in the
same $p_T$ range (see Fig.~\ref{CDF1}).
A possible explanation for the worse agreement at the larger $\sqrt{s}$ values
is that perturbation theory eventually fails if the fraction of available
momentum $z=2p_T\cosh y/\sqrt{s}$ taken away by the produced hadron becomes
too low.

In conclusion, the $p_T$ distribution of inclusive charged-hadron
hadroproduction recently measured by the CDF Collaboration
\cite{Aaltonen:2009ne} at the Tevatron was confronted here for the first time
with rigorous NLO predictions relying on the factorization theorem of QCD,
which, unlike MC event generators, have no left-over adjustable parameters.
Surprisingly, a spectacular disagreement was discovered, by up to 3 orders of
magnitude, at the largest $p_T$ values, where perturbation theory is expected
to work most reliably.
In fact, the $\chi^2$ per degree of freedom turned out to be as large as 7.3
for the last 4 data points.
Perfoming a careful error analysis, we estimated the overall uncertainty
at the largest $p_T$ values to be about $\pm30\%$, being mainly due to
unknown corrections beyond NLO manifesting themselves in residual scale
variations.
In fact, the theoretical uncertainty actually reduces with increasing value of
$p_T$ simply because of asymptotic freedom causing the strong-coupling
constant to fade, so that the high-$p_T$ region is where we would expect the
agreement with the data to be the best.
Being at most of $O(\Lambda_{\rm QCD}/p_T)$, higher-twist effects are bound to
be negligibly small there.
The only large logarithms that need to be resummed are those in
$\alpha_s(\mu)$, the PDFs, and the FFs, which are duly taken care of by the
renormalization group and evolution equations, respectively.
The partonic cross sections are devoid of large logarithms requiring
resummation, which is ensured by the scale choice $\mu=\mu_f=O(p_T)$.
Because $\langle z \rangle \approx 0.5$, no large logarithms of the types
$\ln z$ or $\ln (1-z)$ need to be resummed, and resummation may not even be
appropriate.
Because even the highest $p_T$ bin is far away from the edge of phase space,
no high-$p_T$ resummation is required either.
A different philosophy of error analysis may lead to a somewhat different
result, but certainly not to an error of $O(10^5 \%)$, required to remove the
disagreement.
The latter is even far too large to be remedied by introducing new heavy
virtual particles with resonating propagators or charged particles in the final
state.
Some more drastic modification either to the theoretical preduction or to the
experimental data is required.
We do not claim that the discrepancy discovered by us actually challenges the
factorization theorem, nor that the data is wrong, nor that new physics is at
work.
We do not even wish to speculate about possible problems in the experimental
analysis, such as the suppression of the cosmic-ray background, etc.
We only state that there is a discrepancy, whatever its reason might be.
We recall that no such discrepancy was observed in the inclusive single-jet
production data of the same experiment \cite{Aaltonen:2008eq} at considerably
higher $p_T$ values.
High-$p_T$ data from the LHC is needed to verify whether this apparent
final-state factorization breaking is a genuine phenomenon.
We note that the recent inclusive charged-hadron production data from the
ATLAS and CMS Collaborations \cite{Khachatryan:2010xs} at the LHC are limited
to the region
$p_T < 20$~GeV.
Unfortunately, they are presented as yields differential in $p_T$ and cannot
be converted to cross section distributions because the conversion factors are
not presently known.



This work was supported in part by BMBF Grant No.\ 05H09GUE, by DFG Grant No.\
KN~365/5--3, and by HGF Grant No.\ HA~101.

{\it Note added:} We were informed by the CDF Collaboration that they are
currently reanalyzing their published data \cite{Aaltonen:2009ne}.
In the meantime, four more papers \cite{Arleo:2010kw} on the same subject have
appeared.

\end{document}